\def\Ob{\Omega_b}

\def\Ol{\Omega_\Lambda}
\def\Om{\Omega_m}

\documentclass[structabstract]{aa}

\usepackage{txfonts}
\usepackage{graphicx}
\usepackage{epsf}

\begin{document}

\title{Constraints on CDM cosmology from galaxy power spectrum, CMB and 
SNIa evolution}

\titlerunning{Cosmological constraints after WMAP5 }

\author{
L.D. Ferramacho
\inst{1,}\inst{3} 
\and 
A. Blanchard
\inst{1} 
\and 
Y. Zolnierowski
\inst{2} 
}

\institute{Laboratoire d'Astrophysique de  Toulouse-Tarbes, Universit\'e de Toulouse,  CNRS, 14, Avenue E. Belin, F-31400 Toulouse, France\\
 \email{alain.blanchard@ast.obs-mip.fr}
 \and  LAPP, Universit\'e de Savoie, CNRS/IN2P3, Annecy-le-Vieux, France\\
\email{yves.zolnierowski@lapp.in2p3.fr}
 \and CENTRA, Departamento de F\'isica, Ed\'ificio Ci\^encia, Instituto Superior T\'ecnico, Av. Rovisco Pais 1, 1049-001, Lisboa, Porugal  \\
\email{luis.ferramacho@ist.utl.pt}}

\date {Received / Accepted }

\abstract
 {}
   {We examine the constraints that can currently be obtained on standard cold dark matter models from the most currently used data set: CMB anisotropies, type Ia supernovae and the SDSS luminous red galaxies. We also examine how these constraints are widened when the equation of state parameter $w$ and the curvature parameter $\Omega_k$ are left as free parameters. Finally, we investigate the impact on these constraints of a possible form of 
evolution in SNIa intrinsic luminosity. }
   {We obtained our results from MCMC analysis using the full likelihood of each data set.}
   { For the $\Lambda$CDM model, our 'vanilla' model, cosmological parameters are tightly constrained and consistent with current estimates from various methods. When the dark energy parameter $w$ is free we find that the constraints remain mostly unchanged, i.e. changes are smaller than 
the 1 sigma uncertainties. Similarly, relaxing the assumption of a flat universe
leads to nearly identical constraints on the dark energy density parameter 
of the universe $\Omega_\Lambda $ , baryon density of the universe $\Omega_b $,
 the optical depth $\tau$,  the index of the power spectrum of primordial 
fluctuations $n_S$,  with most one sigma uncertainties better than 5\%. More significant changes appear on other parameters: while preferred 
values are almost unchanged, uncertainties for the physical dark matter density $\Omega_ch^2$, Hubble constant $H_0$ and $\sigma_8$ are  typically twice as large.  The constraint on the age of the Universe, which is very accurate for the vanilla model, is 
the most degraded. We found that  different methodological approaches on large scale structure estimates lead to appreciable differences in preferred values  and uncertainty widths. We found that possible evolution in SNIa intrinsic luminosity does not alter these constraints by much, except for $w$, for which the uncertainty is twice as large. 
At the same time, this possible evolution is severely constrained. }
   {  We conclude that  systematic uncertainties for some estimated quantities are similar or larger than statistical ones.
}


\maketitle

\section{Introduction}

The field of cosmology has made impressive developments in the past decade. 
It is now widely accepted that the present universe is well represented by 
a dark energy dominated universe with dark matter being in the form 
of a cold component and matter fluctuations arising from primordial adiabatic 
Gaussian fluctuations generated in the early universe, a picture consistent with simple inflationary models.
 
Such a paradigm was obtained after a remarkable advance in the quality of observational data. The Hubble diagram derived from type Ia supernovae allowed to obtain the first evidence for dark energy and provided measurements on the properties of this component (\cite{Perlmutter}; \cite{Riess98}). Another important contribution came from the results of the Wilkinson 
Microwave Anisotropy Probe (WMAP) (\cite{WMAP1}; \cite{WMAP3}), which have 
provided strong constraints on many cosmological parameters and allowed for precise estimations of cosmological parameters%, namely on the curvature parameter $\Omega_k$ and baryon density $\Omega_{b}$
. 
Those constraints are consistent with a wide set of constraints coming from 
different cosmological tests and have considerably reinforced early claims 
of  flat cosmology (\cite{1997AA...322..365L}; \cite{1998ApJ...509L..65W}). The use of combined constraints is therefore a way to 
increase the accuracy of cosmological parameter estimations and reduce degeneracies.\\
\indent Among the many different cosmological tests available beyond the CMB, two are particularly robust and based on geometrical considerations: the already mentioned Hubble diagram of type Ia supernovae,  and the recent large 
 galaxy surveys that provided a first detection of the predicted baryonic acoustic oscillations (\cite {eisenstein}, E05 hereafter; \cite{Tegmark}, T06 hereafter; \cite{Percival}). All these observations reinforce the case of the  concordance model in cosmology, in which  the Universe appears to be in an accelerated expansion due to some ``dark energy'' whose true nature remains one of cosmology's most compelling mysteries. The ability of the concordance model 
to accommodate recent high precision data is indeed impressive. Although alternative views exist, it is fair to say that the so called concordance cosmology has led to specific {\em predictions} that were verified a posteriori. The most common interpretations for this ``dark energy'' are an unknown energy component of the Universe or a modification in the equations of general relativity (for some reviews on this subject, see Frieman et al. 2008, Padmanabhan 2006). In order to distinguish between  models beyond the $\Lambda$CDM model, one needs to have very good constraints on the equation of state of dark energy by combining reliable results from quality observations. Some multi-observational constraints have been published recently (\cite{Tegmark}; \cite{WMAP5}; Kowalski et al. 2008; \cite{wright2007}; \cite{xia2008}), but do not  include a detailed likelihood  analysis  of the three most recent observational data sets (i.e. information from one or several data sets was reduced to a single number).\\

In this paper, we propose to constrain several cosmological parameters, including the $w$ parameter in the equation of state of dark energy (supposed constant), using the recent results on the matter power spectrum of LRG galaxies, SNIa and WMAP-5 year data. Although there are other promising methods to test cosmological models and dark energy properties, such as galaxy clusters (\cite{vikhlinin}, \cite{Allen}) or weak lensing (\cite{benjamin}), we have not included
 these in our analysis due to the level of incertitude that may still be 
 present in these methods (\cite{Ferramacho}; \cite{Vauclair}; \cite{Nusa}; \cite{2005AA...436..411B} ). Given the high number of parameters to be constrained, we adopt the Markov chain Monte Carlo (MCMC) method to perform our analysis. The paper is structured as follows: In section 2 we present and discuss the quality of the data. In Section 3 we describe the method and the cosmological models to be constrained and in Section 4 we present the obtained results. Our conclusions are given in Section 5.

\begin{figure*}[t!]
%\centerline{\psfig{file=plots_1D_WMAP5_unionsn_2.eps,width=0.99\textwidth}}
\centering
\hspace*{-17mm}
\includegraphics[width=1.1\textwidth]{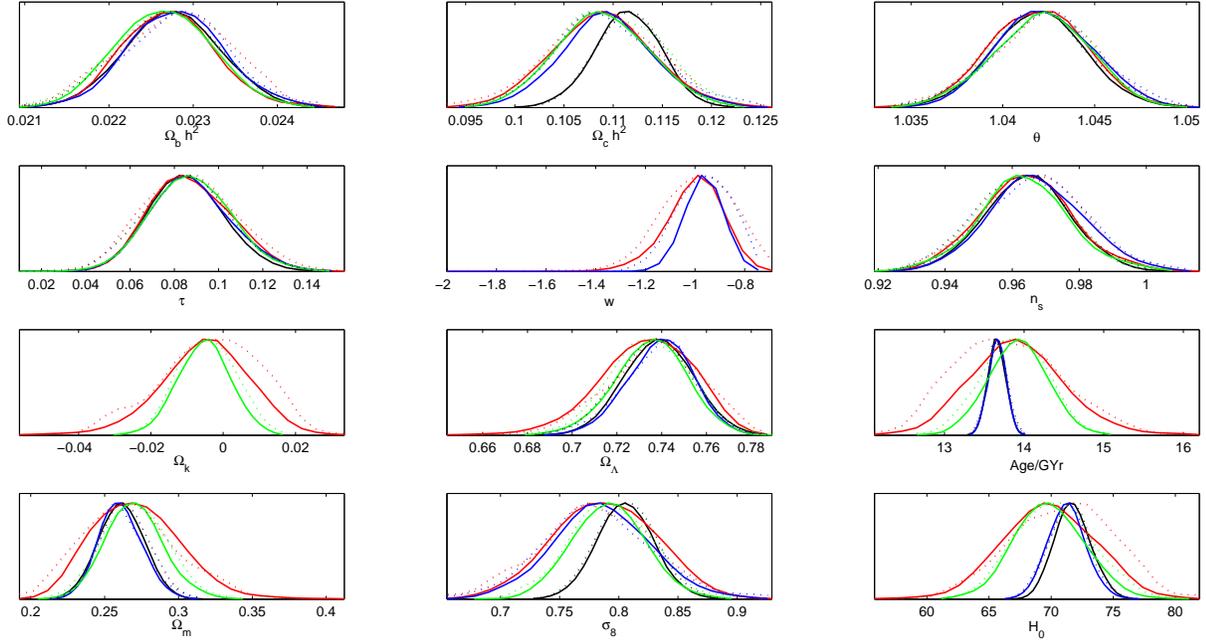}
\vspace*{-35mm}
\caption[]{Posterior distributions for the parameters constrained in our MCMC analysis. Doted curves represent the mean likelihood of the samples, while solid lines show the fully marginalized posterior. The four parameter sets constrained are represented by different colours: Vanilla (black), Vanilla + $\Omega_k$(green), Vanilla + $w$ (blue), Vanilla + $\Omega_k + w$ (red). 
\label{1Dcontour}
}
\end{figure*}

\begin{figure*}[tcb!]
%\centerline{\psfig{file=plots_contours_WMAP5_unionsn.eps,width=\textwidth}}
\includegraphics[width=0.93\textwidth]{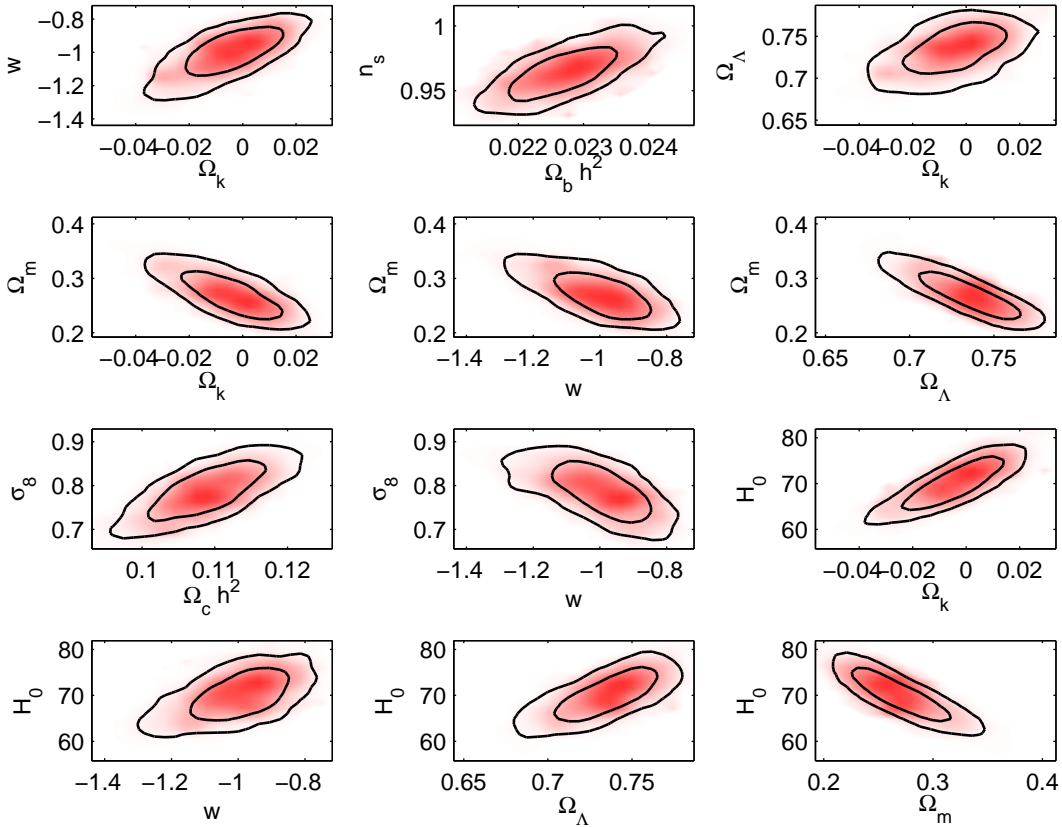}
\vspace*{-8mm}
\caption[]{2D constraints on most correlated cosmological parameters when Vanilla+$\Omega_k+w$ is considered. The curves show the marginalised 1$\sigma$ and 2$\sigma$ confidence regions}
\label{2Dcontour}
\end{figure*}

 \begin{table*}[t!]
%\begin{minipage}{170mm}

\centering
\begin{tabular}{c c c c c l}
\hline
$\mathbf{Parameter}$		&$\mathbf{Vanilla}$	 &$\mathbf{Vanilla + \Omega_k}$			&$\mathbf{Vanilla+w}$ & $\mathbf{Vanilla+\Omega_k+w}$  & \bf{Parameter Definition}\\
\hline
%\multicolumn{4}{l}{Parameters directly constrained:}&\\
%\cline{1-5}
$        \Ob h^2$      &$ 0.0227\pm 0.0005$                            &   $0.0227\pm0.0006$           & $0.0228\pm0.0006$      & $0.0227\pm0.0005$ 	        &Normalised baryon matter density times $h^2$\\
$           \Omega_c h^2$      &$0.112\pm0.003 $                     &  $0.109\pm0.005$            & $0.109\pm0.005$      & $0.109\pm0.005$	        &Normalised CDM density times $h^2$\\
$                 \theta$      &$ 1.042\pm0.003$                       &  $1.042\pm0.003$            & $1.042\pm0.003$	    & $1.042\pm0.003$	 	&Ratio of sound horizon to angular diameter distance\\
$                 \tau$      &$ 0.085\pm0.017$                       &  $0.088\pm0.017$            & $0.087\pm0.017$	    & $0.088\pm0.017$	        &Reionization optical depth\\
$             n_s$      &$0.963\pm0.012$                             &  $0.964\pm0.013$            & $0.967\pm0.014$	    & $0.964\pm0.014$	        &Primordial spectral index at $k=0.05 \rm Mpc^{-1}$\\
$                        log(10^{10} A_s)$      &$3.07\pm0.04$        &  $3.06\pm0.04$            & $3.06\pm0.04$	    & $3.06\pm0.04$	        &$A_s$ is the primordial scalar power at $k=0.05 \rm Mpc^{-1}$ \\
$                       \Omega_k$      &$ 0$                         &  $-0.005\pm0.007$            &  $ 0 $	    & $-0.005\pm0.0121$	        &Spatial curvature   \\
$                        w$      &$ -1 $                             &  $ -1$            &  $-0.965\pm0.056$	    & $-1.003\pm0.102$	        &Dark energy EoS ($w=p_\Lambda/\rho_\Lambda$)\\
\hline
%\multicolumn{4}{|l|}{Derived  parameters:}&\\
%cline{1-5}
$                      \Ol$      &$ 0.738\pm0.015                $   &  $0.735\pm0.016$           & $0.739\pm0.014$      & $0.733\pm0.020$		&Dark energy density ($h^{-2}\rho_{\Lambda}$)\\
$                      Age       $      &$  13.7\pm0.1$              &  $13.9\pm0.4$            & $13.7\pm0.1$	    & $13.9\pm0.6$	        &Age of the universe in Gyr\\
$                      \Om$      &$ 0.262\pm0.015$                   &  $0.270\pm0.019$            & $0.261\pm0.020$	    & $0.272\pm0.029$		&Normalised matter density\\
$                    \sigma_8$      &$0.806\pm0.023$                 &  $0.791\pm0.030$            & $0.816\pm0.014$      & $0.788\pm0.042$		&Matter density fluctuation amplitude \\
$                   z_{re}$      &$10.9\pm1.4$                       &  $11.0\pm1.5$            & $11.0\pm1.5$	    & $11.0\pm1.4$		&Reionization redshift\\
$                    h $      &$ 0.716\pm0.014$                      &  $0.699\pm0.028$            & $0.713\pm0.015$	    & $0.698\pm0.037$		&Hubble parameter\\

\hline
\end{tabular}
%\end{minipage}
\caption{Summary of the posterior distribution mean values for the different sets of parameters constrained, with the corresponding 68\% confidence intervals.}
\end{table*}

\section{Data}

\subsection{CMB}

\indent We use the 5-year data from the WMAP team available at website {\sf http://lambda.gsfc.nasa.gov}, which consists of the all-sky anisotropy maps of 
the cosmic microwave background (CMB) in fits format 
together with a likelihood code which allows us to read these
 maps and calculate the
 TT and TE power spectra and use them to obtain the likelihood of a given 
cosmological model. We also include the recent  ACBAR data (\cite{ACBAR}) which provide additional 
stringent data on small scales.  The new generation of codes publicly available
that compute very efficiently the predicted CMB anisotropies in the line of CMBFAST (\cite{cmbfast}) have 
revolutionised the approach of constraining cosmological parameters from CMB data.

\subsection{SDSS LRG galaxies}

\indent The shape of the present day matter power spectrum depends on  conditions that prevailed during inflation (or more generally during very early stages of the universe) producing the initial fluctuations and on the 
detailed contents of fields and matter of the universe which alter the growth of fluctuations during the expansion.
%  {\it cut of the universe} {\it and which cut} 
This complex process is summarised through the transfer function. In  simple inflationary cold dark matter models, the initial spectrum is  assumed to follow a power law :
$ P(k) = Ak^n$ and  the final power spectrum can be determined using the transfer function which is specified as soon as the cosmological parameters are set, including 
neutrino properties. These transfer functions can be obtained through public codes like CMBFAST (\cite{cmbfast}) or CAMB (\cite{CAMB}). The distribution function of galaxies on large scales in the present day universe is supposed to be representative of the present distribution
of dark matter providing that one can specify a biasing scheme for galaxy formation.
The SDSS survey (\cite{sdss}) allowed to obtain unprecedented data on the large scale distribution of galaxies. In this paper we use the power spectrum published in T06 and derived from 58,380 Luminous Red Galaxies (LRG) along with narrow window functions and uncorrelated minimum-variance errors. For comparison purposes, we will also use the data from E05 on the correlation function derived from 46,748 LRG, with the covariance matrix between data points. Both these data sets can thus be used for likelihood computation, after providing a model which takes into account various corrective effects, such as non-linear evolution and redshift-space distortions.

\subsection{SNIa}

Since the first results that showed evidence for an accelerated expansion of the Universe using a relatively small number of distant supernovae, the number of supernovae available for determination of cosmological constraints has significantly increased due to several supernova observation programs. \cite {Kowal} have compiled the most up to date set of supernovae including the recent large samples of SNIa from SNLS  (\cite{Astier}) and ESSENCE (\cite{essenceI}) surveys, older data sets and the recently extended data set of distant supernovae observed with HST. The compilation contains 414 SNIa and after selection cuts reduces to 307 SN1a. In this paper we use the supernovae magnitude and covariance data provided by \cite {Kowal}  on the 307 SN1a which passed selection cuts.

\section {MCMC analysis and results}

The constraints were obtained using the public code COSMOMC (\cite{lewis}), which
implements a Metropolis-Hastings algorithm to perform a Markov chain Monte Carlo (MCMC) using the above data. This method is the one most currently used; we do not address the question
of the difference between Bayesian and frequentist approaches, which could lead to differences in preferred values and uncertainty estimates (see for instance \cite{dunkley}). We
investigated four different sets of parameters to be constrain by this method, the simplest being
our ``vanilla'' model with $\mathbf{p}$=($\Omega_c h^2$, $\Omega_b h^2$, $\theta$, $\tau$, $n_s$, $A_s$). The $\theta$ parameter represents the ratio between the sound horizon and the angular diameter distance to the last scattering surface. It is used instead of the Hubble parameter $h$ since $\theta$  provides a better efficiency and less correlation with other parameters. The physical meaning of all constrained parameters is described in Table 1. 
By imposing a flat Universe with $w=-1$, it is possible to describe all the used  data with just these six parameters.
These parameters were directly constrained by COSMOMC, while other important
parameters as $h$ or $\Ol$ were derived from those six basic parameters. The two parameters required  to correct for non-linearity and bias effects (\cite{Tegmark})  were introduced in the code that computes the SDSS likelihood, but marginalised analytically.
The other parameter sets consisted of adding the dark energy equation of state parameter $w$ and the curvature parameter $\Omega_k$ to our vanilla model. 

In order to obtain a better and faster convergence, we ran a first chain up to around 10\ 000 points and then used the results to obtain a good estimate of the covariance matrix for the parameters.
This matrix was then used as an input to perform a final chain run until convergence.  

We ran one chain for each parameter set until it satisfied the Raftery-Lewis convergence test (\cite{RafteryLewis}) 
which typically happened after 15\ 000-30\ 000 accepted iterations. The posterior distribution was then used to derive the confidence intervals for the parameters. A flat prior was attributed to all directly constrained parameters, and 
{ there is} a restriction on $h$ ($0.4<h<1.0$) imposed in CAMB. Table 1 shows the marginalised mean values for all parameters constrained, either directly or indirectly, as well as the 1$\sigma$ confidence intervals. We note the fact that the quoted values do not represent exactly the best fit model in either case, but the mean and standard deviation of the marginalised probability distribution of each parameter. These distributions are shown in Fig.1 for comparison between the different parameter sets. We also present the 2D marginalised distributions for the main cosmological parameters in the case where 8 parameters were directly constrained (Fig. \ref{2Dcontour}). This figure allows us to identify possible 
correlations among parameters. 
  
\begin{figure}[!h]
%\centerline{\psfig{file=plots_contours_WMAP5_unionsn.eps,width=\textwidth}}
\includegraphics[width=0.4\textwidth]{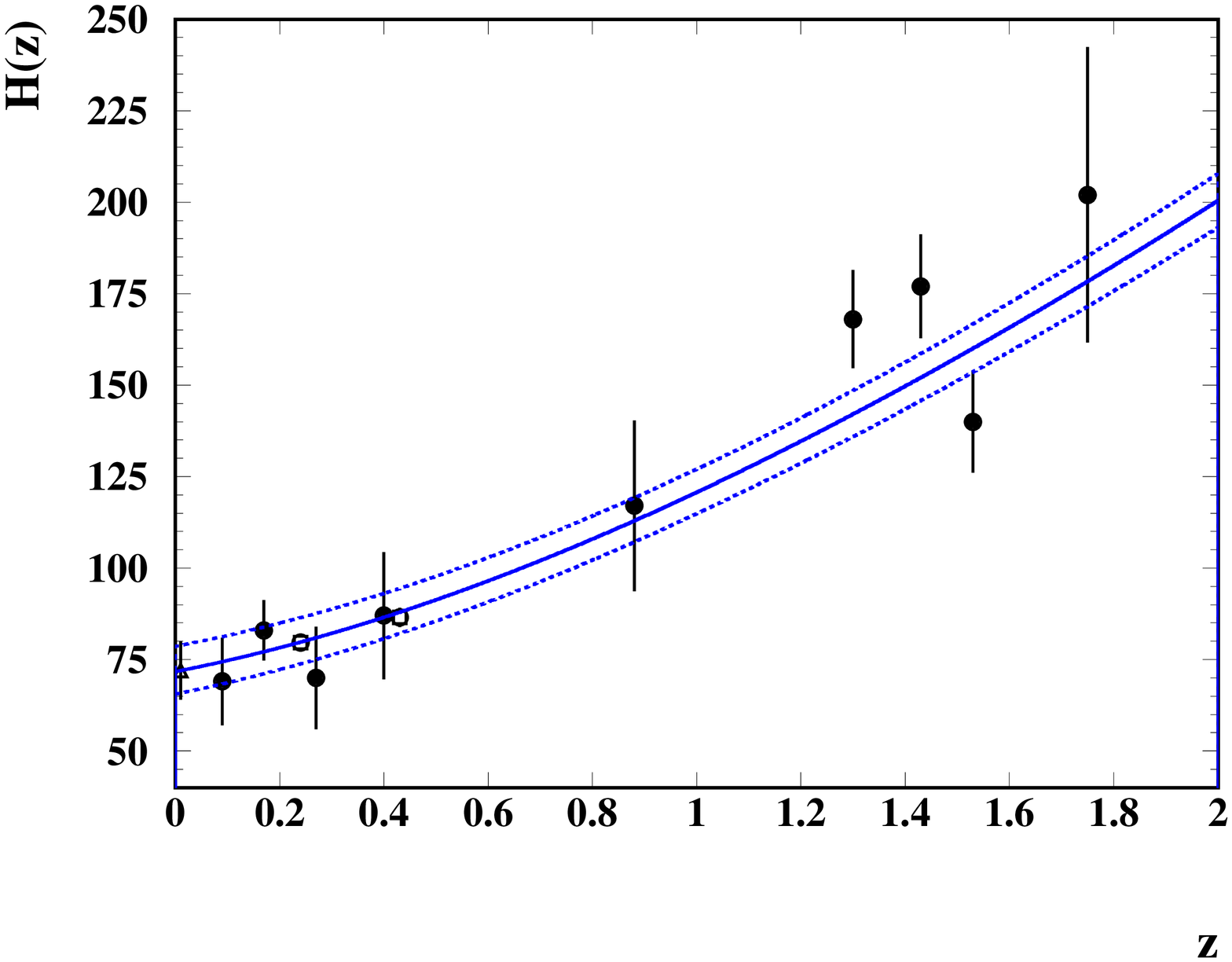}
\caption[]{Comparison between the data on the Hubble parameter evolution from 
three data set (triangle: \cite{HubbleHST}; dark circles \cite{H(z)}; open circles : \cite{Gasta08}) and one $\sigma$ envelop from our constraints on $\Lambda CDM$ vanilla model.}
\label{Hz}
\end{figure}

Finally, we compared the data published   on the evolution of the Hubble parameter $H(z)$ (\cite{H(z)}; \cite{Gasta08}) with the region allowed by our constraints on the vanilla model (figure \ref{Hz}). This was done by taking the envelop  of $H(z)$ 
of models falling inside 1 $\sigma$ confidence intervals in the distributions for $\Omega_m$ $\Omega_{\Lambda}$, $\Omega_c h^2$. This comparison illustrates 
the ability of the $\Lambda$CDM model to fit accurately various type of astrophysical data, although the evaluation of possible systematics is difficult for such type of data.

%the random points from the posterior distributions of  $\Omega_{\Lambda}$, $\Omega_c h^2$, $\Omega_b h^2$ and selecting the sets of values which fell inside the
%1 $\sigma$ confidence intervals in the 2D distributions for $\Omega_m$ $\Omega_{\Lambda}$,  $\Omega_c h^2$, $\Omega_b h^2$. For each value of $z$, we have retain the minimum and maximum values of $H(z)$. 

\section{Systematics}

\subsection{LSS analysis and compressed data}

\begin{figure*}[tcb!]
%\centerline{\psfig{file=plots_1D_WMAP5_unionsn_2.eps,width=0.99\textwidth}}
\hspace*{6mm}
\includegraphics[width=0.9\textwidth]{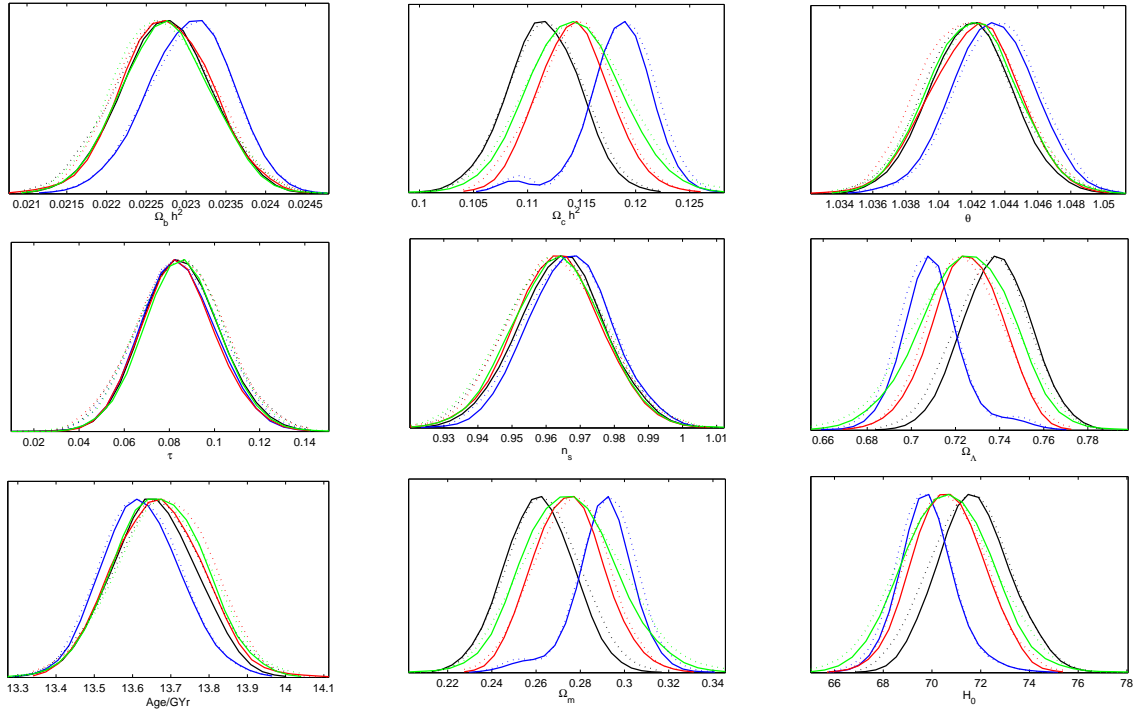}
\caption[]{Joint marginalised posterior distributions for relevant cosmological parameters in a vanilla model comparing four approaches to the LSS data: Matter power spectrum and corrections for non-linear effects from T06 (\emph{black}), correlation function from E05 with corrections for non-linear effects from E05 (\emph{red}) and T06 (\emph{blue}). The final constraint uses the distance parameter $A$ derived in E05 (\emph{green}).
\label{1Dcontour_comp}
}
\end{figure*}

Table 1 provides an up to date estimation of cosmological parameters by 
combining the likelihood from each data set. Clearly, the accuracy is impressive.
Given this precision, it is vital to evaluate possible systematics that could seriously alter these estimations. 

 One possible source of systematic uncertainty lies in the analysis of the large scale galaxy %mass
 distribution. %Most of the current 
Most of recent analyses include SDSS data on BAO. %are based on the papers of 
%(\cite{eisenstein}, \cite{Tegmark})
 In E05 and T06, the authors have used different descriptors to quantify the galaxy clustering. In principle, the use of the correlation function or power spectrum should give the same results, but
%the fact of 
using different numerical estimators can lead to different answers, providing an indication on the amplitude of possible biases.
%introduce some level of bias between both approaches that should be tested for.
 Another sensible aspect is that of non-linear corrections. In order to compare the observed galaxy clustering with that from underlying dark matter, one has to account for changes in the linear power spectrum (or correlation function) due to non-linear gravitational collapse and scale dependent bias of galaxies. Such effects introduce distortions on the linear power spectrum shape and can even shift the position of the acoustic peak of up to 3\% (\cite{corBAOnl}). Several models have been proposed to correct for these effects, usually supported by numerical simulations. In T06, non-linear corrections were performed using a one parameter model introduced by \cite{Cole}:
%{\it S. Cole et al., MNRAS, 362, 505 (2005)}
%
\begin{equation}
P_{gal}=P_{dewigg} b^2\frac{1+Q_{nl}k^2}{1+1.4k}.
\label{Qmodel}
\end{equation}

Here $P_{dewigg}$ stands for the linear power spectrum corrected for a power suppression of the baryonic oscillations due to the displacement of galaxies induced by their particular velocities. The other part of the above expression quantifies the scale dependent distortion due to non-linear gravity and bias.  

In E05, the authors used the same correction for the baryonic peak supression, but the scale dependent bias and non-linear collapse were treated differently, by means of the HALOFIT formalism and an expression derived from the numerical simulation of \cite{Seo2008} and applied directly to the real space correlation function.

We have thus compared 
%investigated then the effect of including 
the different approaches referred to above, when combined with the CMB and SNIa information. We restricted this analysis to the vanilla model and the results are shown in Fig. 3 for relevant parameters. One can see that using the E05 two point correlation function and non-linear corrections results in a difference
%discrepancy 
in the mean value of some parameters compared to using the T06 data on the matter power spectrum together with a Q-model. The parameters most affected  are the densities $\Omega_c h^2$, $\Om$, $\Ol$ and the Hubble parameter $H_0$. For instance, $\Om$ is shifted from $0.262$ to $0.275$ while the uncertainty is of the order of $0.015$, the difference is therefore of the order of 1 $\sigma$ or more. 

In order to %determine the main contribution 
track the origin of such a difference, we ran another chain using the correlation function data from E05, but with a Q-model  correction for non-linear effects. One could expect that by using the same modelling as T06, the observed difference would be reduced. However, the blue curves in Fig. 3 show that the difference increases in this case, with $\Om=0.295\pm.015$. This result is the same if we leave the $Q_{nl}$ parameter in eq.\ref{Qmodel} as a free parameter or if we use the best value found by T06, $Q_{nl}=31$. Such a result indicates a significant difference between the estimations  in E05 and T06,
% to derive their data points, 
since they both used similar galaxy catalogs. Although the observed differences are no more than 1 to 2 $\sigma$, they are significative enough to consider that the present LSS and BAO data has to be used with some caution. Clearly, future works on this subject must address carefully the issue of bias in estimators (\cite{Norberg2008}) and  modelling of non-linear effects. With this caveat, we stress that the use of the matter power spectrum was less time consuming and we kept this approach for wider parameter investigations.      
      
It is also interesting to notice that several
recent papers provide constraints from combination of 
these three data sets, but in which some of the constraints were obtained 
from highly compressed information, typically one number, for 
instance in the form of a 
reduced distance. Obviously different compressed information from the same data could result in 
different constraints (see for instance Lewis, 2008); the use of compressed 
information and approximate likelihoods should therefore be 
tested to ensure they do not lead to inadequate 
approximation of 
the likelihood on the full data set. Although some compression is necessary to 
reduce the numerical cost of likelihood analyses, too strongly compressed information
 may result in 
biases on both the preferred values and on the uncertainties. 
 We have used the distance parameter $A$ from E05 to perform our constraints, with the results also plotted in Fig. 3 (red curves). This parameter seems to capture quite accurately the cosmological information from the full correlation function with non-linear corrections used in E05. The mean values for all parameters are highly compatible in both analyses. The main difference lies in the uncertainties of some parameters, which seem to be overestimated by an factor of 25\% to 30\% when we use this reduced distance.

\subsection{Possible astrophysical troubles }

Even in the presence of ideal data with no systematics or in the data acquisition either 
in the data analysis, inferred constraints may be biased because of unidentified complex astrophysics. For instance 
CMB fluctuations are contaminated by various foregrounds, which may introduce bias in estimation if 
incorrectly subtracted (\cite{SZcon}). Similarly, the acoustic scale as well as the whole correlation function inferred from galaxy samples could be biased by
 distortion in redshift space, non-linear dynamics and the biasing of the galaxy population. All
these effects are believed to be controlled to a much higher precision than present surveys allow by now (Seo et al, 2008).

\subsubsection{SNIa evolution and consequences}

The situation for SNIa is more problematic: SNIa are complex objects probably resulting from mass accretion of
a white dwarf in a  binary system; the properties 
of the final explosion may depend on many of the parameters of the progenitors (age, metalicities, environment, ...) and the properties of the explosion are not well known (detonation/deflagration, physics of the ignition points). The SNIa 
results heavily rely on the assumption that the peak luminosity, once corrected for duration--luminosity dependence (\cite{Phillips}), does not vary with 
redshift. Some 
evolution in the progenitor population is expected from the general stellar population. Some  evolution may therefore be present and hard to detect. Indeed, this is one of the main weakness of SNIa results (\cite{RiessLivio}). From 
the observational point of view only differences in the spectroscopic properties of distant and close supernovae can be quantified (Balland et al., 2007; \cite{Foley}). The role of possible grey dust (\cite{2002AA...384....1G}) or evolving dust may also be important  (Holwerda, 2008). 

\begin{figure}[!h]%\centerline{\psfig{file=plots_contours_unionsn_evol.eps,width=\textwidth}}
\includegraphics[width=0.4\textwidth]{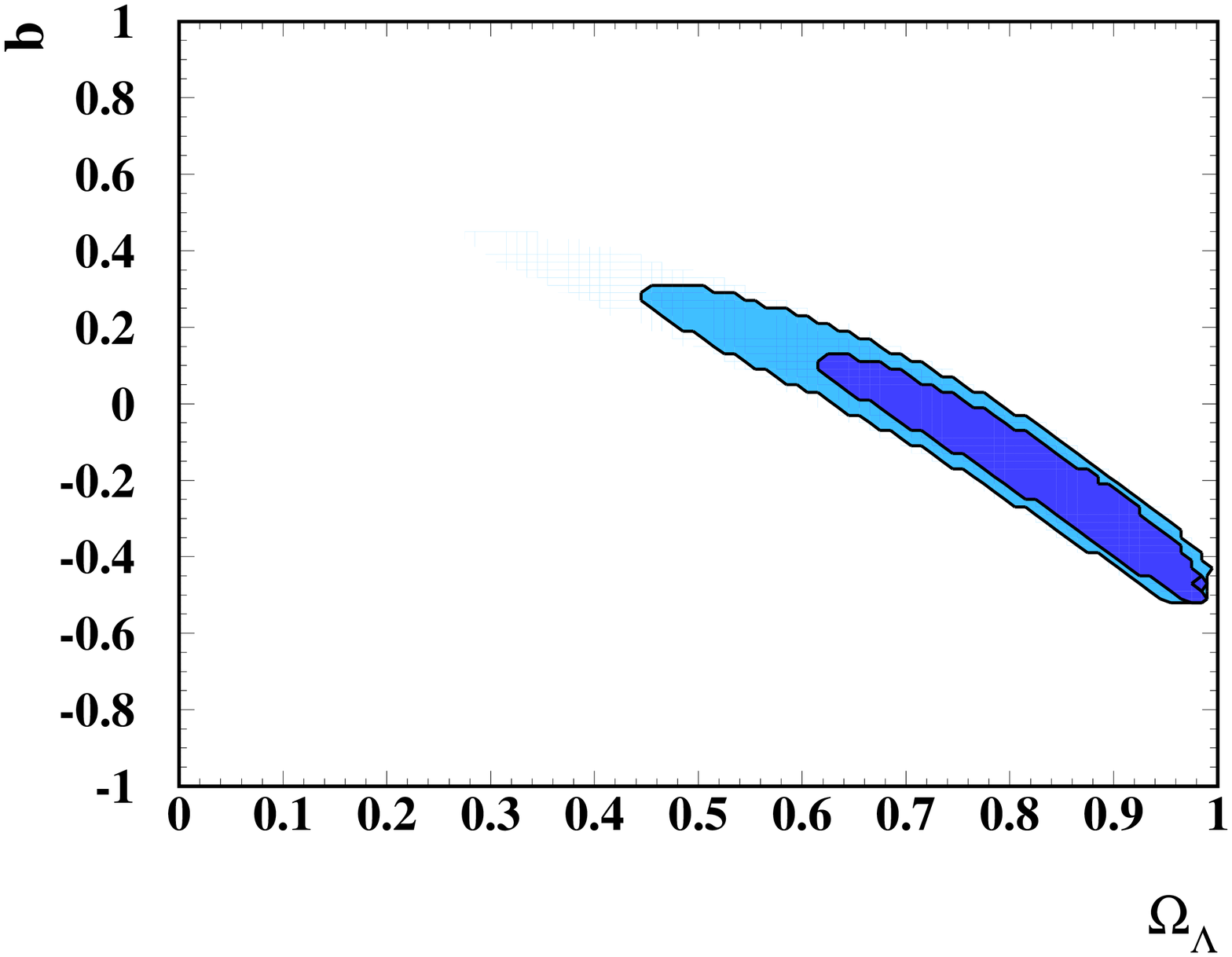}
\includegraphics[width=0.4\textwidth]{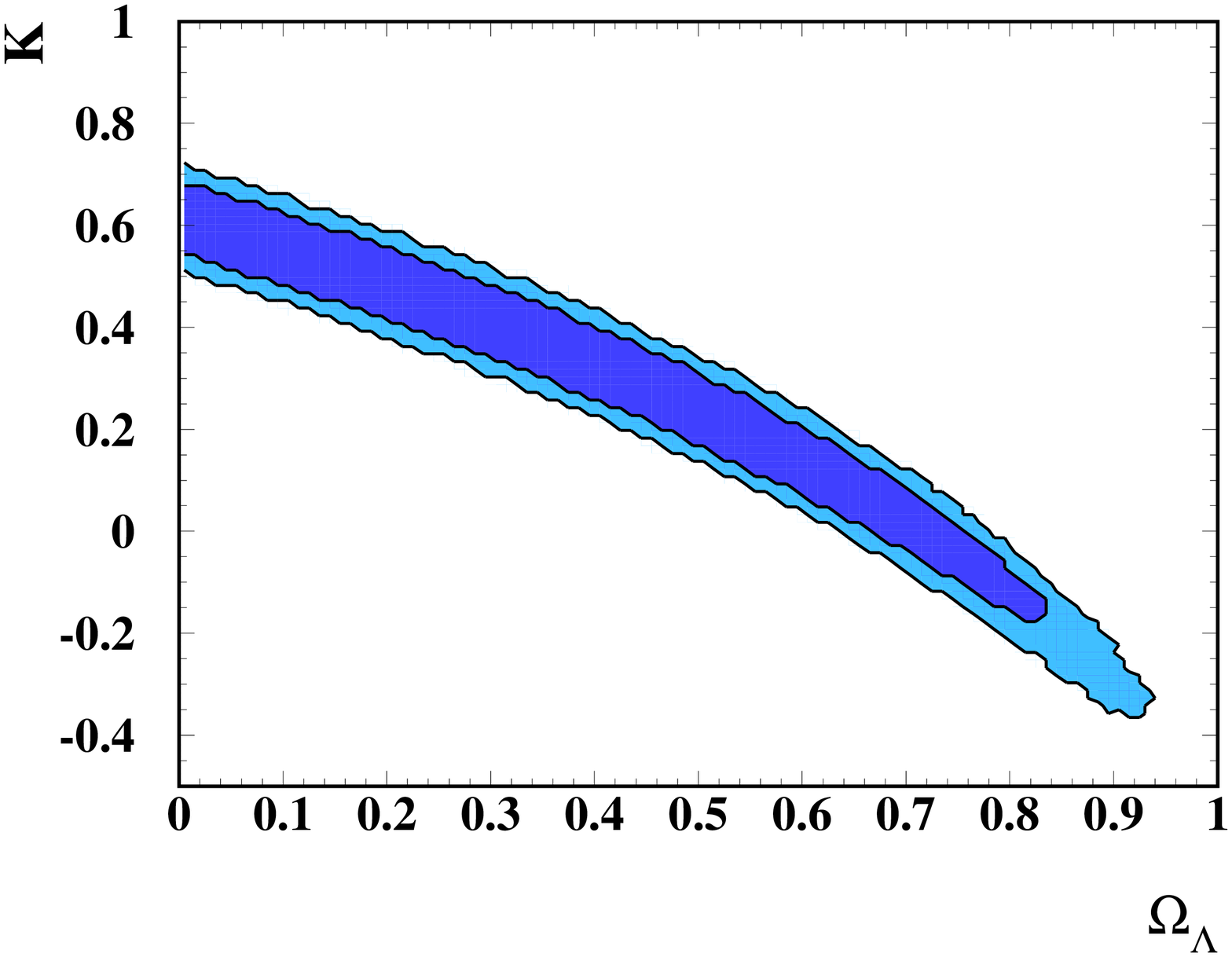}
\caption[]{Constraints in the $\Omega_\Lambda-b$ (top) and $\Omega_\Lambda-K$ (bottom) planes from 
the Hubble diagram for a flat cosmological model. Contours correspond to 1 and 2 $\sigma$ regions on two parameters. }
\label{SNEvol}
\end{figure}

 \begin{table*}[t!]
%\begin{minipage}{170mm}

\centering
\begin{tabular}{c c c c c}
\hline
$\mathbf{Parameter}$		&$\mathbf{Vanilla}$	 &$\mathbf{Vanilla + \Omega_k}$			&$\mathbf{Vanilla+w}$ & $\mathbf{Vanilla+\Omega_k+w}$ \\
\hline
%\multicolumn{4}{l}{Parameters directly constrained:}&\\
%\cline{1-5}
$        \Ob h^2$      &$ 0.0228\pm 0.0006$                            &   $0.0227\pm0.0005$           & $0.0227\pm0.0006$      & $0.0226\pm0.0006$ 	\\        
$           \Omega_c h^2$      &$0.110\pm0.004 $                     &  $0.109\pm0.005$            & $0.113\pm0.005$      & $0.111\pm0.005$	 \\
$                 \theta$      &$ 1.042\pm0.003$                       &  $1.042\pm0.003$            & $1.042\pm0.003$	    & $1.042\pm0.003$	 \\
$                 \tau$      &$ 0.088\pm0.017$                       &  $0.087\pm0.017$            & $0.085\pm0.017$	    & $0.085\pm0.016$	   \\  
$             n_s$      &$0.968\pm0.013$                             &  $0.965\pm0.013$            & $0.963\pm0.014$	    & $0.960\pm0.014$	     \\  
$                        log(10^{10} A_s)$      &$3.07\pm0.04$        &  $3.06\pm0.04$            & $3.07\pm0.04$	    & $3.06\pm0.04$	       \\
$                       \Omega_k$      &$ 0$                         &  $-0.002\pm0.007$            &  $ 0 $	    & $-0.017\pm0.013$	         \\
$                        w$      &$ -1 $                             &  $ -1$                     &  $-1.112\pm0.148$	    & $-1.33\pm0.242$	       \\
$                        K$      &$ -0.042\pm0.042 $                &  $ -0.035\pm0.042$            &  $-0.105\pm0.091$	    & $-0.133\pm0.077$	       \\
\hline
%\multicolumn{4}{|l|}{Derived  parameters:}&\\
%cline{1-5}
$                      \Ol$      &$ 0.747\pm0.017                $   &  $0.745\pm0.020$           & $0.756\pm0.022$      & $0.744\pm0.022$	\\
$                      Age       $      &$  13.6\pm0.1$              &  $13.7\pm0.4$            & $13.6\pm0.1$	    & $14.5\pm0.7$	      \\
$                      \Om$      &$ 0.253\pm0.017$                   &  $0.257\pm0.025$            & $0.244\pm0.022$	    & $0.272\pm0.029$ \\
$                    \sigma_8$      &$0.801\pm0.026$                 &  $0.794\pm0.029$            & $0.846\pm0.068$      & $0.867\pm0.060$	 \\
$                   z_{re}$      &$11.1\pm1.5$                       &  $11.0\pm1.4$            & $10.9\pm1.5$	    & $10.8\pm1.4$\\
$                    h $      &$ 0.725\pm0.017$                      &  $0.720\pm0.036$            & $0.748\pm0.038$	    & $0.703\pm0.042$	\\

\hline
\end{tabular}
%\end{minipage}
\caption{Similar to Table 1, but including an evolution parameter for the supernova intrinsic luminosity.}
\end{table*}

Despite the remarkable efforts on observations of distant SNIa, it is clear that 
the hypothesis of negligible evolution is extremely difficult to prove. As it is commonly written, the absence of evidence is not an evidence of absence.  A simple evolution law could be:
\begin{equation}
\Delta m(z) = b z
\label{lum_evol1}
\end{equation}

\begin{figure}[!h]
%\centerline{\psfig{file=plots_contours_WMAP5_unionsn.eps,width=\textwidth}}
%\hspace{-1 cm}
\includegraphics[width=0.5\textwidth]{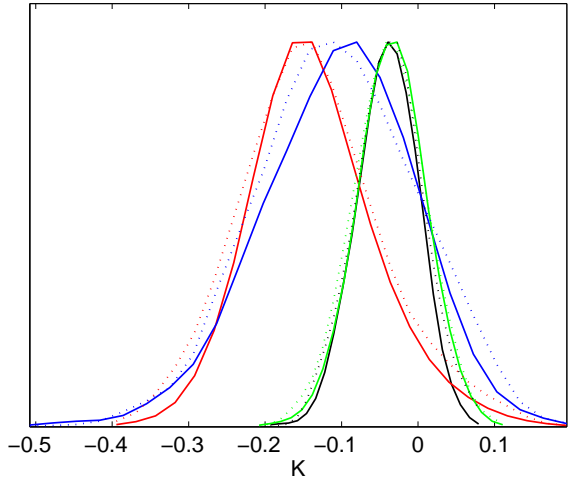}
\caption[]{Posterior distribution for the SN evolution parameter K, for different parameter sets of CDM models. The colour code is the same as the one used in Fig. 1. Evolution is systematically preferred, although the non evolving solution remains acceptable.}
\label{1DcontourEvol}
\end{figure}

(\cite{2008JCAP...02..008N}). Wright (2002) showed that 
an evolution model in  which peak luminosity varies as an exponential function of cosmic time within an Einstein de Sitter universe may 
mimic the accelerated expansion. We have therefore also examined 
the possible consequences on cosmological constraints assuming that the peak magnitude of SNIa
evolved  linearly with time as:
\begin{equation}
\Delta m(z) = K \left(\frac{t_0-t(z)}{t_0-t_1}\right)
\label{lum_evol2}
\end{equation}
where $t_0$ is the present age of the universe,  $t_1$ is the age of the universe at $z = 1$ and  $t(z)$ is the age of the universe at the redshift of the supernova explosion. The parameters $b$ or $K$ represent the typical change in magnitude at redshift 1. We first examined
 the consequence of such evolution for the SNIa Hubble diagram. In Figure \ref{SNEvol} we determined the constraints in the  $\Omega_\Lambda-b$ and $\Omega_\Lambda-K$ planes from 
the Hubble diagram for a flat cosmological model. As one can see, there is a 
degeneracy between the two parameters and no  tight constraint can be obtained on $\Omega_\Lambda$ any longer.  As the second evolution law introduced stronger levels of degeneracies, we concentrated on this second form. We then re-examined the various constraints established in section 3  with our additional parameter describing SNIa evolution.  
This was first done 
on the vanilla model with $K$ as an additional parameter. We found that the cosmological constraints
are almost unchanged. This is not surprising  as we checked that removing SNIa does not change the 
constraint on the vanilla model by much. Our analysis allows us to constraint the amount of possible evolution in SNIa in our model consistent with existing data (CMB and correlation function).
 We found that $K$ is essentially uncorrelated  with the  other parameters. The posterior distribution on $K$ is given in figure \ref{1DcontourEvol}, for our parameter sets. The constraint on $K$ can be summarised by:  
$$ K = -0.042 \pm 0.042$$
for the vanilla model. We also examine how constraints are relaxed when   the curvature parameter, $\Omega_k$ and $w$ are free. Again, the constraints are not modified significantly. Also, there is a clear degeneracy between $w$ and $K$, which degrades constraints
on both parameters. Resulting constraints are summarised in Table 2. For the most general model (curvature and $w$ free)
the constraint on $K$ is three times larger:
$$ K = -0.133 \pm 0.077$$
Our analysis shows that evolution is not required by the data. Most noticeably,  it changes significantly the constraints on $w$:  while the vanilla model is almost at the center of the 1 $\sigma$ contour when evolution is not allowed (Fig. \ref{2DcontourEvol}),  the constraints appear twice as bad (darker contours in figure \ref{2DcontourEvol}) when evolution is allowed, and the best model then lies close  
to the boundary of the 2 $\sigma$ contour without evolution. 

Interestingly enough, constraints on others parameters are quite stable even in the most general case (curvature and $w$ free) and therefore insensitive to such a possible evolution. 
Our results put a stringent constraint on the possible evolution of SNIa
and thereby provide an idea of the level at which systematic uncertainty for 
possible evolution should be controlled in order to provide

\noindent useful constraints
 on the problem of the determination of cosmological parameters in conjunction with other methods: even in the worse case the amount of evolution 
is found to be of the order of $\Delta m \sim -0.136$ (1 $\sigma$ level),
 at redshift $z = 0.5$, while in the vanilla model we found that  evolution is limited to $\Delta m \sim -0.055$ at the same redshift. We notice that {\em spectral} SNIa evolution is constrained to be less than 10\% from the observed properties of a sample of SNIa with a median redshift of 0.5 (\cite{Foley}).

\begin{figure}[!h]
%\centerline{\psfig{file=plots_contours_WMAP5_unionsn.eps,width=\textwidth}}
\includegraphics[width=0.5\textwidth]{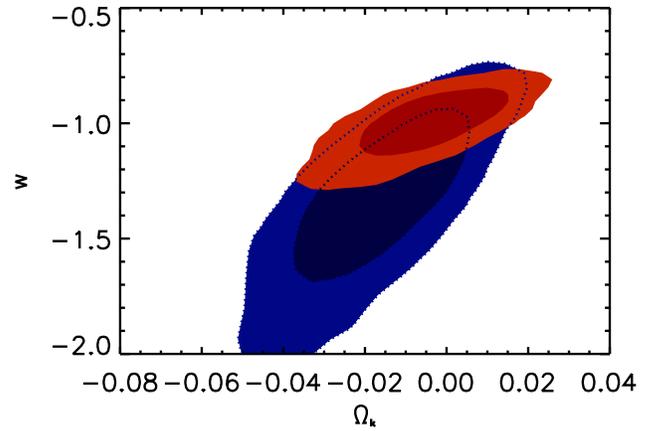}
\caption[]{Joint 2D marginalised constraint on the dark energy equation of
state parameter $w$ (supposed constant) and the curvature density
$\Omega_k$, assuming evolution in the SNIa luminosity (\emph{blue}) and
considering no evolution (\emph{red}). The contours show the confidence
regions of 1 (darker) and 2 $\sigma$. }
\label{2DcontourEvol}
\end{figure}
\begin{figure}[!h]
%\centerline{\psfig{file=plots_contours_WMAP5_unionsn.eps,width=\textwidth}}
\includegraphics[width=0.5\textwidth]{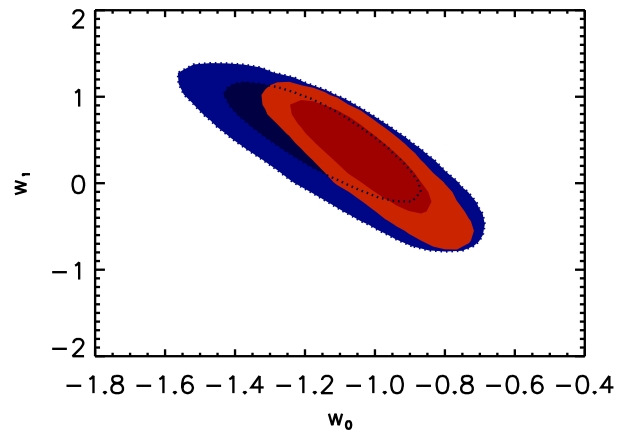}
\caption[]{Joint 2D marginalised constraint on the dark energy equation of
state parameter $w$ (supposed constant) }
\label{2DcontourEvol}
\end{figure}

 Finally, we have examined how the so-called figure of merit (f.o.m), defined 
according to the \cite{DETF}  (see for instance \cite{VireyEalet}), is modified when  possible  evolution in the SNIa of the above type is taken into account. We therefore used the CPL formulation (\cite{CPL1}; \cite{CPL2}) for dark energy evolution. The state equation parameter is supposed to evolve according to:
\begin{equation}
\centering
w(z) = w_0+(a-1)w_1
\label{wCPL}
\end{equation}
where $a$ is the expansion factor normalised to one at the present epoch. 
We restricted ourselves to the flat case. The f.o.m is defined
as the area of the 90\% contour in the $w_0,w_1$ diagram. The joint constraint contours are presented in figure \ref{2DcontourEvol}.
The f.o.m. is found to be 2.03 when no  evolution was assumed and is lowered to 1.17 when evolution is considered. This is in  agreement with our previous finding~: taking into account evolution leads to degraded constraints although the loss in precision is limited.   

\section{Discussion and conclusions}

In this paper, we have used three of the most solid sets of observational data in 
cosmology to perform an up to date constraint on the $\Lambda$CDM model: CMB, the 
Hubble diagram of distant SN and the shape of the power spectrum of galaxy 
distribution on large scales. We have chosen to use the power spectrum shape 
because the analysis using the 
correlation function was appreciably longer.  We have confirmed that 
  6 parameters only are enough 
 to reproduce the three currently used data sets. Using appropriate likelihoods for each data set,  we obtained very good 
constraints with typical  uncertainties of the order of a few percent. The comparison between the values obtained in this work
 and those found  using only WMAP3+SDSS (\cite{Tegmark})  data do not show 
significant changes in most of the parameters, which demonstrates the great 
constraining power of these large scale observations. Indeed, the introduction 
of the supernova data produces changes mostly on the parameters that 
govern the geometry of the Universe, namely $\Ol$ and $h$, which have their 
mean values reduced by around 0.02/0.03. As for the dark energy EoS parameter 
$w$, we do not obtain a better constraint by adding the supernova data, and the results point to a value highly compatible with a cosmological constant although the 1$\sigma$ error bars are appreciably larger than on other parameters (of the order of 5 to 10 \%) .          

We found that adding extra freedom to the vanilla model does not 
decrease the quality of the constraints in an appreciable way for several parameters:  as can be seen from figure \ref{1Dcontour}, constraints on the baryonic content of the universe $\Omega_b$, the dark energy content $\Ol$, the matter content of the universe $\Omega_c$ and the primordial index of scalar fluctuations $n_S$ are almost unchanged.  We get a slightly wider distribution in other  parameters,
with preferred values unchanged but with uncertainties increased by a factor of up to two.
%  although the difference is tiny in a number of  significant parameters (content of the universe and the amplitude of matter fluctuations),  
The most drastic change happens for the age of the universe, with an uncertainty increased by a factor of more than five, although the uncertainty remains still only of the order of 5\%.

Our constraints have been established within some assumptions. For instance 
we do not investigate the possible contribution of tensor modes (Zaldarriaga \& Seljak 1997; Liddle \& Lyth 2000) or neutrino mass (\cite{2006JCAP...06..019G}) . 
Obviously, the addition of new freedom could modify some of our constraints.
There is no general method to avoid this type of limitation; for instance,
one of the most critical assumptions on constraints obtained from the CMB is
the power law shape of the spectrum of primordial fluctuations. Relaxing this assumption could lead to dramatic modification of the constraints (\cite{2003AA...412...35B}, \cite{2007PhRvD..76l3504H}). 
%No big changes are seen in the mean values obtained with the different parameter sets, except for $\Omega_c h^2$, where we see a notable decrease when both $\Omega_k$ and $w$ are included in the analysis. This may be due to...  

While preparing this paper, we benefited from the release of the 5-year WMAP results. The WMAP team presents MCMC constrains on the $\Lambda CDM $ model (\cite{WMAP5}), using WMAP+SN+BAO. They used the BAO measurements on the angular diameter scale $D_A$ derived from the SDSS LRG correlation function (\cite{eisenstein}), while we preferentially used the full power spectrum shape data. 
By comparing the two approaches
%our results 
%with those from Komatsu et al. (2008), we can see that for the 6-parameter case,  
we find very similar constraints on all parameters 
and similar 
%agree very well and are compatible within the 
$1\sigma$ confidence intervals, which indicates that the 
%two approaches are equivalent. 
use of the distance parameter $A$ in implementing the LSS constraint is essentially sufficient.  
%This is not surprising since the CMB analysis including only the 5-year WMAP data already provide stringent constraints  (\cite{dunkley}) on cosmological parameters close to our final  constraints.
Also, we find that the use of the power spectrum instead of the correlation function  leads to appreciable difference of up to two sigma.
Finally our investigation on the consequence of a specific model for the evolution of SNIa luminosities has been interesting in several respects: first, although evolution does not allow any conclusion on cosmological parameters from the Hubble diagram alone, we found that most of the constraints from combinations remained
 unchanged, and at the same time the amount of evolution is severely 
constrained to be less than 5 to 20\% for redshifts between 0.5 to 1.

The current status of cosmology seems then to be highly compatible with the $\Lambda$CDM. Present-day existing data allows to estimate cosmological parameters to within a few percent whithin a specific model. This raises the fundamental question of the role of astrophysical sources of uncertainties.
Clearly the next step is to understand in more detail the nature of dark energy (cosmological constant, dynamic scalar field or modified gravity) and this is yet to be achieved. The ever increasing quality of data, anticipated from future large observational programs, could be used to rule out or validate some of theses alternatives. However, future  projects would need to keep systematic effects, both from methodological and astrophysical origin, controlled to achieve a typical anticipated precision of one percent or better.

%Our future work should pass by using this data to constrain some alternative models for dark energy.           

\begin{acknowledgements}

L.Ferramacho acknowledges financial support provided by Funda\c c\~ao para a Ci\^encia e Tecnologia (FCT, Portugal) under fellowship contract SFRH/BD/16416/2004.  

\end{acknowledgements}

\end{document}